\documentclass[twoside]{article}
\usepackage{fleqn,espcrc2}
\usepackage{graphicx}

\renewcommand{\Im}{\mathop{\rm Im}}

\title{\vbox to 0pt{\vskip-1in
\rightline{\normalsize Preprint MPI-PhT/2000-02}}\vskip-8pt
Effects of CP-Violation in Neutralino Scattering and Annihilation}

\author{Paolo Gondolo \address{Max-Planck-Institut f\"{u}r Physik,
    F\"{o}hringer Ring 6,
    80805 M\"{u}nchen, Germany (gondolo@mppmu.mpg.de)} 
  and Katherine Freese$^{\rm ~a,}\!\!$
  \address{Physics Department, University of Michigan, Ann Arbor, MI 48109,
USA (ktfreese@umich.edu)}
}

\begin{document}

\maketitle

We show that in some regions of supersymmetric parameter space, CP violating
effects that mix the CP-even and CP-odd Higgs bosons can enhance the neutralino
annihilation rate, and hence the indirect detection rate of neutralino dark
matter, by factors of $10^6$. The same CP violating effects can reduce the
neutralino scattering rate off nucleons, and hence the direct detection rate of
neutralino dark matter, by factors of $10^{-7}$. We study the dependence of
these effects on the phase of the trilinear coupling $A$, and find cases in the
region being probed by dark matter searches which are experimentally excluded
when CP is conserved but are allowed when CP is violated.  

The neutralino elastic scattering cross section (in pb) is plotted in fig.~1 as
a function of neutralino mass (in GeV) for $\sim 10^6$ values in SUSY parameter
space.  The upper panel is for the case of CP violation via $\Im(A) \neq 0$
while the lower panel is for the case of no CP violation.  In the upper panel,
it is the maximally enhanced cross section (as a function of $\arg(A)$) that is
plotted.  The dark points refer to those values of parameter space which have
the maximum value of the cross section for nonzero $\Im(A)$ and which are
experimentally excluded at zero $\Im(A)$. The grey region refer to those values
of parameter space which are enhanced when CP violation is included and which
are allowed also at zero $\Im(A)$.  The light grey empty squares refer to those
values of parameter space which have no enhancement when CP violation is
included.  The solid lines indicate the current experimental bounds placed by
DAMA and CDMS; the dashed lines indicate the future reach of the CDMS (Soudan),
GENIUS, 
\includegraphics[width=6.6cm]{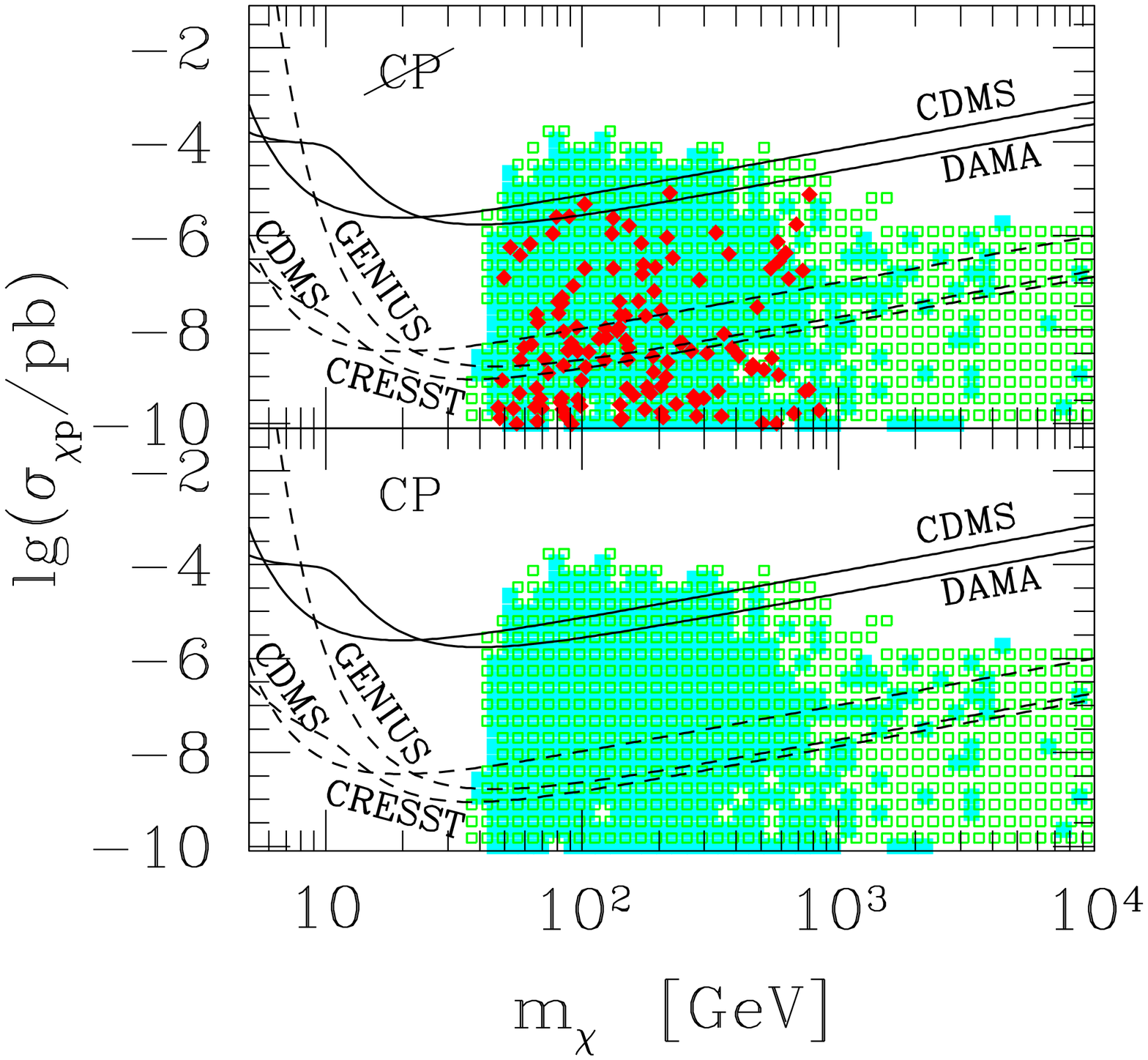} 

\vspace{-2\baselineskip} {\tt Figure 1.} 
\vspace{1.2\baselineskip}

\noindent 
and CRESST proposals.

In fig.~2 we show the enhancement and suppression factors of the elastic
scattering cross section for the case of CP violating $\arg(A)$.  The plot
shows the ratios $R_{\rm max} = \sigma^{\rm max}/\max[\sigma(0),\sigma(\pi)] >
1$ and $R_{\rm min} = \sigma^{\rm min}/\min[\sigma(0),\sigma(\pi)]$ as a
function of the values $\phi_A$ of the phase of $A$ where the maximum/minimum
occur Here $\sigma^{\rm max}$ ($\sigma^{\rm min}$) is the enhanced (suppressed)
scattering cross section and the superscript max (min) indicates the maximal
enhancement (suppression) as one goes through the phase of $A$.  The
denominator of the ratio $R_{\rm max}$ ($R_{\rm min}$) chooses the larger
(smaller) value of the scattering cross section without CP violation, i.e., for
phase = 0 or phase = $\pi$.

Fig.~3 shows the enhancement and suppression factors of the neutralino
annihilation cross section times relative velocity $\sigma v$ (at $v=0$). The
ratios $R^{\rm ann}_{\rm max}$ and $R^{\rm ann}_{\rm min}$ are defined
similarly to $R_{\rm max}$ and 
\includegraphics[width=6.6cm]{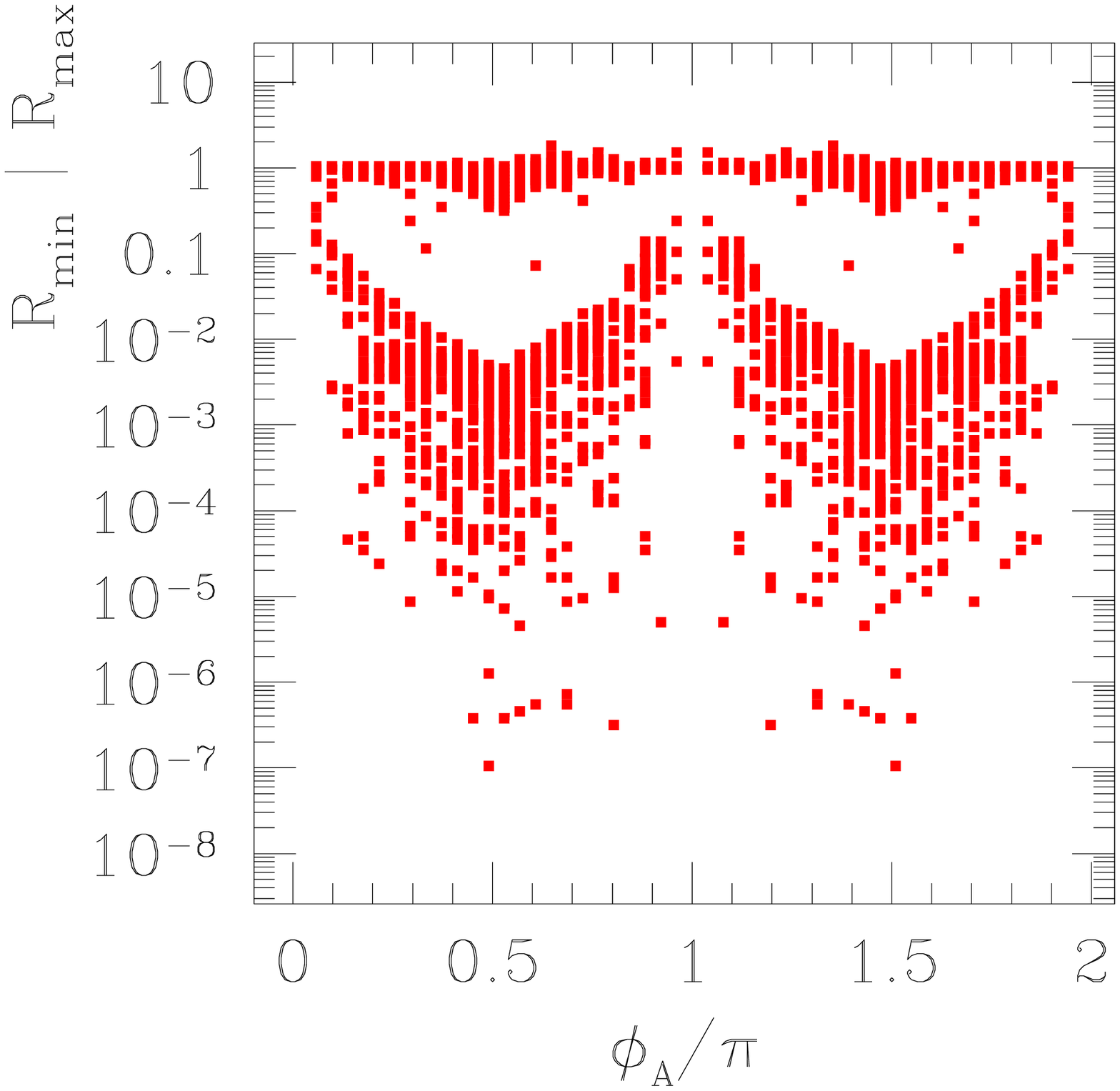} 

\vspace{-2\baselineskip}  {\tt Figure 2.} 
\vspace{1.22\baselineskip}

\noindent 
$R_{\rm min}$ but with $\sigma v$ replacing the
scattering cross section $\sigma$.

Finally, in figs.~4 and 5 we show two examples of the behavior of the
scattering and annihilation cross sections with the phase of $A$.  The four
panels from top to bottom display the following: the scattering cross section
$\sigma_{\chi p}$ in pb, the annihilation cross section $\sigma v$ in cm$^3$/s,
the branching ratio $\mathop{\rm BR}(b\to s\gamma) \times 10^4$, and the
lightest Higgs boson mass $m_{h_1}$ in GeV as a function of the phase $\phi_A$
of $A$. CP conserving phases are $\phi_A = 0, \pi$ while all other values are
CP violating.  In the third and fourth panels we hatch the regions currently
ruled out by accelerator experiments.  In all four panels we denote the part of
the curves that is experimentally allowed by thickened solid lines, and the
part that is experimentally ruled out by thinner solid lines.  In this figure,
the possible phases are bound by the limit on the $b\to s\gamma$ branching
ratio. In the allowed regions, the scattering cross section at CP-violating
phases is suppressed, while the annihilation cross section is enhanced. The
latter takes its maximum allowed value when the $b\to s\gamma$ limit is
reached.  In the case plotted in fig.~4, both CP conserving cases are
experimentally excluded while some CP violating cases are allowed.  The
scattering cross section is of the order of $10^{-6}$ pb, and lies in the
region being probed by direct detection experiments. The annihilation cross
section peaks at $\phi_A=3\pi/4$; notice that this value is not the point of
maximal CP violation $\phi_A=\pi/2$.
\includegraphics[width=6.6cm]{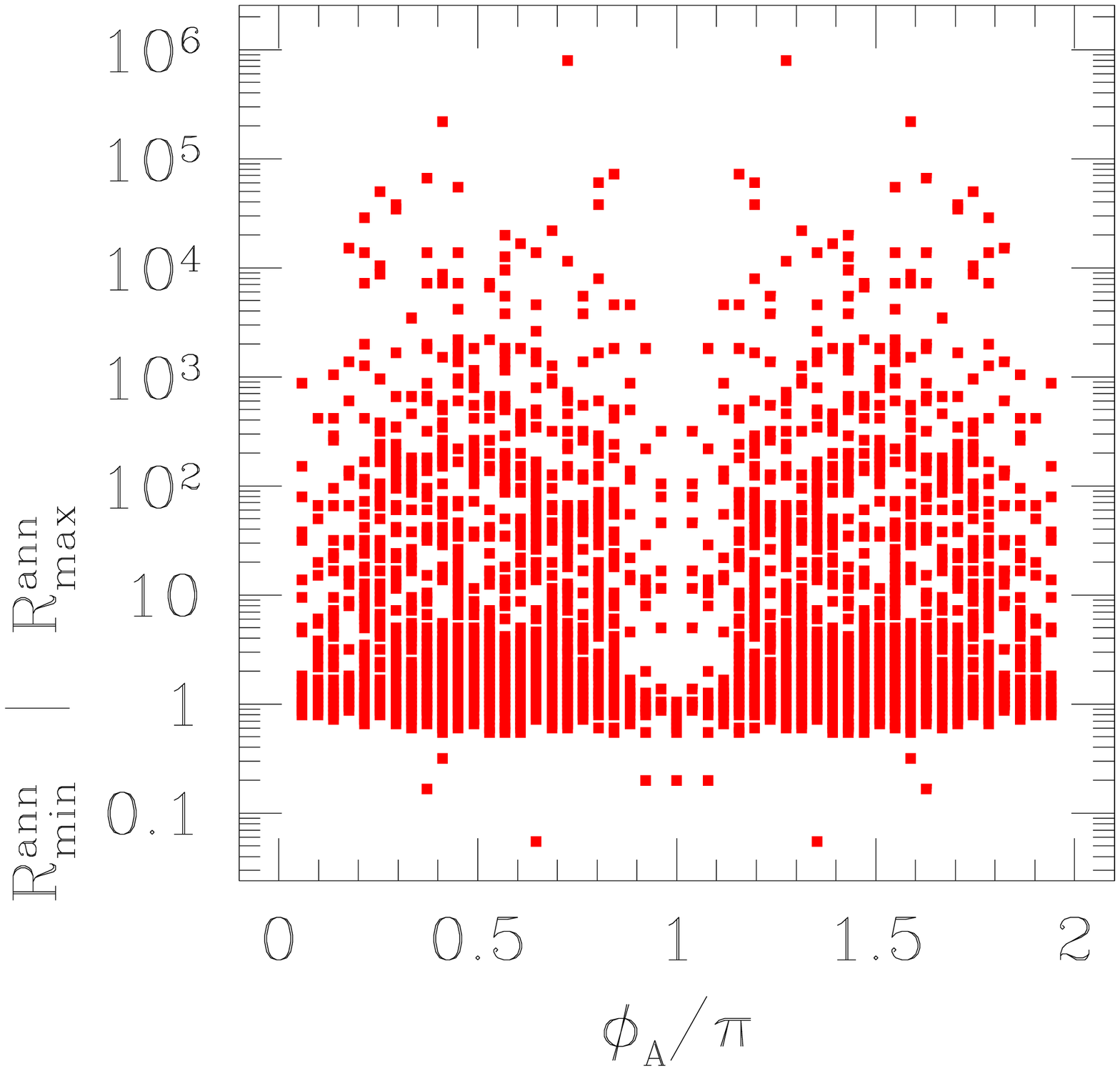}  

\vspace{-2\baselineskip}  {\tt Figure 3.}

\vspace{0.2\baselineskip}

\includegraphics[width=6.6cm]{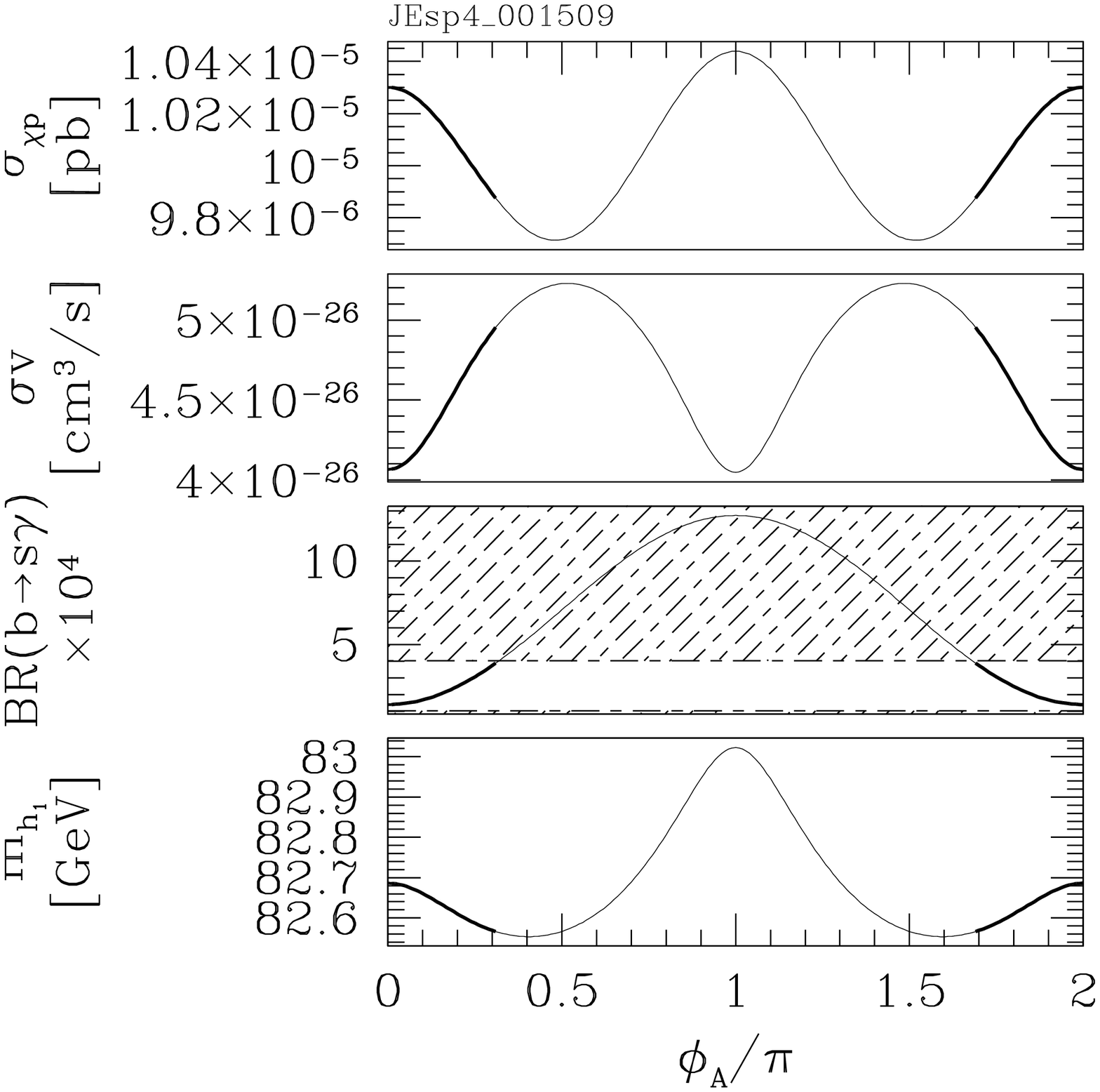}  

\vspace{-2\baselineskip}  {\tt Figure 4.}

\vspace{0.2\baselineskip}

\includegraphics[width=6.6cm]{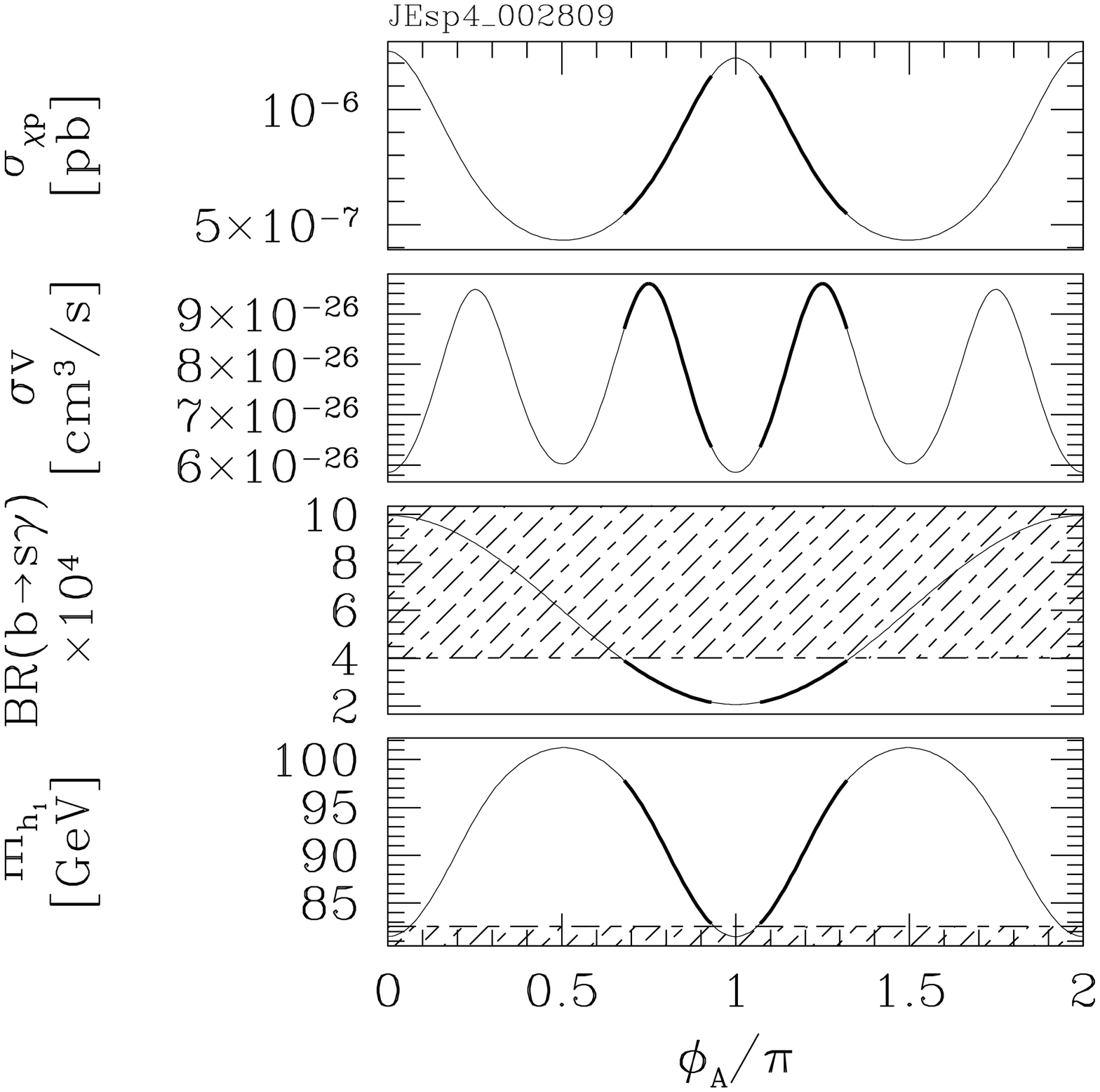}  

\vspace{-2\baselineskip}  {\tt Figure 5.}
\vspace{1\baselineskip}

A detailed presentation and
references can be found in Gondolo and Freese,
hep-ph/9908390.

\end{document}